\documentclass[12pt]{article}
\usepackage{psfig}

\def\refb#1{(\ref{#1})}
\newcommand{\be}{\begin{equation}}
\newcommand{\ee}{\end{equation}}
\newcommand{\ba}{\begin{eqnarray}}
\newcommand{\ea}{\end{eqnarray}}
\def\laq{\raise 0.4ex\hbox{$<$}\kern -0.8em\lower 0.62ex\hbox{$\sim$}}
\def\gaq{\raise 0.4ex\hbox{$>$}\kern -0.7em\lower 0.62ex\hbox{$\sim$}}
\begin{document}
\title{Cosmic Black Holes}
\author{Eun-Joo Ahn\thanks{E-mail: sein@oddjob.uchicago.edu}\\
\small Department of Astronomy and Astrophysics\\
\small and \\
\small Center for Cosmological Physics \\
\small University of Chicago, 5640 S.~Ellis Ave., Chicago, IL 60637 USA\\
Marco Cavagli\`a\thanks{E-mail: marco.cavaglia@port.ac.uk}\\
\small Institute of Cosmology and Gravitation\\
\small University of Portsmouth, Portsmouth PO1 2EG, United Kingdom}
\date{}
\maketitle
\begin{abstract}

Production of high-energy gravitational objects is a common feature of
gravitational theories. The primordial universe is a natural setting for the
creation of black holes and other nonperturbative gravitational entities.
Cosmic black holes can be used to probe physical properties of the very early
universe which would usually require the knowledge of the theory of quantum
gravity. They may be the only tool to explore thermalisation of the early
universe. Whereas the creation of cosmic black holes was active in the past, it
seems to be negligible at the present epoch.

\end{abstract}

{\it This essay received an ``honorable mention'' in the 2003 Essay Competition
of the Gravity Research Foundation -- Ed. }

\newpage 

The physics of the very early universe is one of the unsolved questions in
cosmology: How did the universe begin? Standard cosmology \cite{Kolb:vq}
speculates that the universe has a history of continuous expansion. If we
believe in the standard cosmological model, the universe began some 10-15
billions of years ago from a singularity in a superhot and superdense state,
where temperature and energy density reached and maybe surpassed Planck values.
The very early universe is quantum and its understanding requires a quantum
theory of gravity. This conclusion is not unique to the standard cosmological
scenario. We know that the theory of general relativity breaks down at high
energies. No matter what may be the theory of quantum gravity, if we define the
Planck scale $M_{\rm Pl}$ as the energy scale where general relativity breaks
down, we may conclude that the early universe requires a Planck scale. Although
the physics of the Planck state depends on the details of the quantum theory,
the conclusion that the universe has passed through a Planckian stage, as
previously defined, cannot be avoided. For instance, standard cosmology and the
``new'' string \cite{Gasperini:1992em} and brane-world \cite{Khoury:2001wf}
cosmologies predict the existence of this phase. However, almost nothing is
known of the Planckian era. We do not even know whether the primordial universe
began in a thermal equilibrium state.

Can we say anything about the very early universe without any knowledge of
Planckian physics? Surprisingly, the answer to this question is in the
affirmative. There is indeed a physical phenomenon which is common to all
theories of gravity and that can be used to gain information about the
Planckian regime of gravity: Black holes (BHs). BH solutions, or more generally
solutions with trapped surfaces, are a common feature of gravitational
theories. The details of the solution depend on the theory under consideration
but the existence of trapped surfaces cannot be avoided. We may hope to obtain
general information on the Planckian era of the universe using general
properties of BHs. In this essay we present a simple, though completely
general, example of this technique.

Apart from gravitational collapse of matter \cite{ZeldovichNovikov66}, BHs may
be created from Planckian particle collisions \cite{Banks:1999gd}. The geometry
of these BHs depends on the geometry of the spacetime. In $D$ dimensions the
simplest BH has spatial spherical topology. Solutions with spatial topology
$R^p\times SO(D-p-1)$ are usually called (black) branes \cite{Duff:1993ye}.
Nonperturbative creation of BHs \cite{Banks:1999gd}, branes
\cite{Ahn:2002mj,Ahn:2002zn}, and other gravitational extended objects is
expected to dominate hard-scattering processes at Planckian scales. (For a
review, see, e.g., Ref.~\cite{Cavaglia:2002si,Tu:2002xs}.) If
the Planck scale happens to be very low, as in gravitational theories with
large extra dimensions, formation and decay of BHs and branes would soon be
detectable in hadron colliders \cite{Giddings:2001bu} and high-energy cosmic
ray detectors \cite{Feng:2001ib}. The very early universe is a natural arena
for Planckian BH production: Creation of nonperturbative gravitational objects
could have been a common event in the early universe. 

Primordial creation of nonperturbative gravitational objects by particle
collision may give some information on the initial conditions of the universe.
We have already mentioned that we do not know whether the universe was in
thermal equilibrium during its earliest epoch. All perturbative interactions
are frozen out at very high energy and are ``[\dots] ineffective in maintaining
or establishing thermal equilibrium'' \cite{Kolb:vq}. In the standard scenario,
the energy threshold for thermalisation is $10^{16}$ GeV, where Grand
Unification Theory processes freeze out. A rough estimate of the thermalisation
threshold for a given process can be obtained by comparing the interaction rate
per particle $\Gamma$ to the expansion rate of the universe:
\be
\Gamma\,\gaq\, H\,.
\label{theq}
\ee
Here $H$ is the Hubble parameter, and $\Gamma=n\sigma\langle|v|\rangle$, where
$n$ is the number density of the species, $\sigma$ is the cross section of the
process, and $v$ is the relative velocity. Assuming a standard
radiation-dominated epoch for the early universe and interactions mediated by
massless gauge bosons, Eq.~\refb{theq} reads
\be
\Gamma/H\sim \alpha^2 T^{-1}\,\gaq\, 1\,,
\label{theq-standard}
\ee
where $T$ is the temperature in Planck units and $\alpha\sim O(10^{-1})$ is the
gauge coupling constant. Therefore, for temperatures smaller than $\sim
10^{16}$ GeV the interactions occur rapidly and the universe thermalises,
whereas at temperatures above $10^{16}$ GeV the interactions cannot maintain
thermal equilibrium. Hence thermal equilibrium of the very early universe
cannot be achieved with perturbative processes.

At temperatures about (and above) the Planck scale the energy per particle is
sufficient to trigger nonperturbative gravitational processes. Therefore, we
expect BHs and other gravitational objects to be in thermal equilibrium with
the primordial bath. Once formed, the BHs accrete and evaporate by quantum
Hawking radiation \cite{Hawking:sw} until they reach equilibrium, thus
thermalising the universe. Thermalisation is possible because BH formation is a
nonperturbative process. Furthermore, thermalisation does not depend on the
details of the process such as the geometry of the BH. At temperatures of order
of the Planck scale we expect the gravitational object to decouple from the
thermal plasma. After decoupling, the remaining BHs either evaporate completely
and disappear or leave some stable remnants \cite{Cavaglia:2003qk}. We will
briefly come back to this point later on. 

Let us make the previous statements quantitative. Although the physics of
nonperturbative gravitational interactions in the primordial universe requires
a careful study, a simple argument shows how gravitational effects may indeed
thermalise the universe. Consider the formation of a gravitational object with
event horizon of size $r_h$. The mass of this object increases with increasing
$r_h$ and increasing spacetime dimension, i.e.,
\be
M\sim r_h^{a}\,,
\label{mass}
\ee
where $a \ge 1$ is a constant parameter which depends on the number of
spacetime dimensions. The formation of the gravitational object can be modelled
as a semiclassical process if the entropy of the process is sufficiently large,
i.e., if the fluctuations of the number of (micro) canonical degrees of freedom
are small \cite{Cavaglia:2002si}. Gravitational objects with  mass equal to a
few Planck masses usually satisfy this condition. In this case, the
cross-section is approximated by the geometrical cross section of the object.
Assuming that the colliding particle lives in a $(k+4)$-dimensional submanifold
of the $D$-dimensional spacetime ($k\le D-4$), the cross section in Planckian
units is
\be
\sigma(s)\sim M^{k+2\over a}\sim s^{{k+2}\over 2a}\,,
\label{cross}
\ee
where $\sqrt{s}$ is the centre-of-mass energy of the colliding particles. For
instance, the cross section of a nonrotating spherically symmetric BH
\be
ds^2=-R(r)dt^2+R(r)^{-1}dr^2+r^2d\Omega^2_{D-2}\,,~~~~~~
R(r)=1-\left({r_{s}\over r}\right)^{D-3}\,,
\label{Schwarzschild}
\ee
is
\be
\sigma_{BH}\sim A_{k,D}\,s^{k+2\over 2(D-3)}\,,
\label{cross-BH}
\ee
where $A_{k,D}$ is a geometrical factor of order $O(1)$. Since we want to
discuss the thermal equilibrium of BHs, we need to write the cross section as a
function of the temperature. This is simply Eq.~\refb{cross} with the
substitution $s\to T^2$:
\be
\sigma(T)\sim T^{2+k\over a}\,.
\label{cross-T}
\ee
As is expected, the cross-section \refb{cross-T} increases as the temperature
increases. BHs thermalise the universe at high temperatures. This statement is
independent from the geometry of the gravitational object and is completely
general. The interaction rate $\Gamma$ is easily estimated assuming for
simplicity that the $(k+4)$-dimensional space is homogeneous and isotropic. In
this case the number density and the interaction rate are  $n\sim T^{k+3}$ and
$\Gamma\sim T^{1+(1+1/a)(k+2)}$, respectively. Comparing the interaction rate
to the the expansion rate $H\sim T^{k/2+2}$, the condition for thermal
equilibrium \refb{theq} reads
\be
T^{(1+2/a)(1+k/2)}\,\gaq\, 1\,.
\ee
Therefore, the universe is thermalised by nonperturbative gravitational
interactions at temperatures $\,\gaq\,M_{\rm Pl}$. If the primordial universe is
Planckian, it is also thermal. When the temperature drops below the Planck
scale, the nonperturbative gravitational interactions freeze out and BH
accretion essentially stops. The ``last'' gravitational objects which have been
produced in the Planckian era are expected to decay and disappear after a time
scale of the order of $M_{\rm Pl}^{-1}$. BHs, for instance, evaporate with
emission of Hawking radiation. However, the possibility that some of the
Planckian gravitational objects are stable and survive as stable relics should
not be discarded \cite{Ahn:2002zn, Cavaglia:2003qk}.

Does cosmic production of gravitational objects play a role at the present
epoch as well? The answer to this question depends obviously on the value of
the fundamental Planck scale. If the Planck scale is as low as a few TeV, it is
in principle possible to observe cosmic creation of BHs and other gravitational
objects. These events could be produced by the collision of a high energy
relativistic particle, e.g.\ an ultra-high energy cosmic ray, with a massive
particle such as a dark matter particle. Since the cross section grows as a
function of the centre-of-mass-energy of the collision, the highest rate of
formation is obtained when the energy of the relativistic particle $E$ is large
and the Planck scale is small. In order to produce a gravitational object, the
mass $m$ of the massive particle must be larger than $M_{\rm Pl}^2/2E$. The
best scenario is obtained by setting $M_{\rm Pl}\sim$ 1 TeV and choosing the
highest-energy cosmic rays with energy $E_\nu\sim 10^{21}$ eV. In this case,
the minimum particle mass required to create gravitational objects is $m\sim 1$
keV. Warm and cold dark matter such as WIMPs \cite{Kolb:vq} would be good
candidates. The number of events, however, decreases drastically with the mass
of the particle \cite{thanks}. The interaction rate in Planckian units is %
\be
\Gamma\,\laq\,\rho_c (2E)^{1/a} m^{1/a-1}\,,
\label{intrate}
\ee
where $\rho_c$ is the critical density. Therefore, the interaction rate
decreases with increasing mass $m$. BHs, for instance, have $a=D-3$ and at
least five dimensions are needed to lower the Planck scale to TeV values. If
cosmic BHs are created today, collision of high-energy cosmic rays with warm
dark matter maximizes BH formation. The interaction length $\lambda$ for cosmic
creation of spherical BHs in Planckian units is given by $\Gamma^{-1}$: 
\be
\lambda\,\gaq\,\rho_c^{-1} (2E)^{-1/a} m^{1-1/a} \,.
\label{lambda}
\ee
Choosing $E=10^9$ $M_{\rm Pl}$, $m=10^{-8}$ $M_{\rm Pl}$, $M_{\rm Pl}=1$ TeV
and five dimensions, we find $\lambda\sim 10^{33}$ cm $\gg$ Hubble distance.
Therefore, cosmic production of BHs is negligible at the present epoch. 

To conclude, cosmic creation of gravitational objects is a general feature of
any gravitational theory. The formation of gravitational objects is a
nonperturbative phenomenon which takes place when the energy of the process is
of the order of the Planck scale or higher. Cosmic creation of BHs and other
objects happens during the primordial phases of the universe, when the
temperature is expected to have reached Planckian values. A striking effect of
the formation of gravitational objects is the thermalisation of the universe at
very high energies, which cannot be achieved by perturbative processes. BH
formation in the very early universe allows to get some information about the
initial conditions of the universe without making any assumption on the quantum
theory of gravity. Cosmic BH formation triggered by particle collisions at very
high energy cannot occur today, even if the Planck scale is as low as a TeV.
However, the merging BH physics and cosmology is a field full of promises and
may be well bring further surprises.  

\vskip 1em
\leftline{\large\bf Acknowledgements}

\noindent

We thank Jos\'e Blanco-Pillado for interesting discussions. M.~C.\ thanks the
University of Chicago for the kind hospitality.

\end{document}